\documentclass[twocolumn]{aastex62}

\graphicspath{{./}{}}

\received{January 1, 2018}
\revised{January 7, 2018}
\accepted{\today}
\submitjournal{ApJ}

\shorttitle{The Sizes of $z\sim 9-10$ galaxies}
\shortauthors{Holwerda et al.}

\begin{document}

\title{The Sizes of $z\sim9-10$ Galaxies Identified in the BoRG Survey}

\correspondingauthor{Benne W. Holwerda}
\email{benne.holwerda@louisville.edu}

\author[0000-0002-4884-6756]{Benne W. Holwerda}
\affiliation{Department of Physics and Astronomy, 102 Natural Science Building, University of Louisville, Louisville KY 40292, USA}

\author[0000-0002-8584-1903]{Joanna S. Bridge}
\affiliation{Department of Physics and Astronomy, 102 Natural Science Building, University of Louisville, Louisville KY 40292, USA}

\author[0000-0001-9537-5814]{Rebecca L. Steele}
\affiliation{Department of Physics and Astronomy, 102 Natural Science Building, University of Louisville, Louisville KY 40292, USA}

\author[0000-0002-0761-1985]{Samir Kusmic}
\affiliation{Department of Physics and Astronomy, 102 Natural Science Building, University of Louisville, Louisville KY 40292, USA}

\author[0000-0002-7908-9284]{Larry Bradley}
\affiliation{Space Telescope Science Institute, 3700 San Martin Drive, Baltimore 20218 MD, USA}

\author{Rachael Livermore}
\affiliation{ARC DECRA Fellow, School of Physics, The University of Melbourne, Parkville, VIC, 3010, Australia}

\author{Stephanie Bernard}
\affiliation{School of Physics, University of Melbourne \& 
ARC Centre of Excellence for All-Sky Astrophysics (CAASTRO), Australia}

\author{Alice Jacques}
\affiliation{Department of Physics and Astronomy, 102 Natural Science Building, University of Louisville, Louisville KY 40292, USA}




\begin{abstract}
Redshift $z=9--10$ object selection is the effective limit of Hubble Space Telescope imaging capability, even when confirmed with Spitzer. If only a few photometry data points are available, it becomes attractive to add criteria based on their morphology in these J- and H-band images. 

One could do so through visual inspection, a size criterion, or alternate morphometrics. We explore a vetted sample of BoRG $z\sim9$ and $z\sim10$ candidate galaxies and the object rejected by Morishita+ (2018) to explore the utility of a size criterion in z=9-10 candidate selection. A stringent, PSF-corrected effective radius criterion ($r_e<0\farcs3$) would result in the rejection of 65-70\% of the interlopers visually rejected by Morishita+. It may also remove up to $\sim20$\% of bona-fide brightest ($L>>L^*$) z=9 or 10 candidates from a BoRG selected sample based on the Mason+ (2015) luminosity functions, assuming the Holwerda+ (2015) $z\sim9$ size-luminosity relation. We argue that including a size constraint in lieu of a visual inspection may serve in wide-field searches for these objects in e.g. EUCLID or HST archival imaging with the understanding that some brightest ($L>>L^*$) candidates may be missed. 

The sizes of the candidates found by Morishita+ (2018) follow the expected size distribution of $z\sim9$ for bright galaxies, consistent with the lognormal in Shibuya+ (2015) and single objects. Two candidates show high star-formation surface density  ($\Sigma_{SFR} > 25 M_\odot/kpc^2$) and all merit further investigation and follow-up observations.  

\end{abstract}


\section{\label{s:intro}Introduction}

Near-infrared deep observations with the \emph{Hubble} and \emph{Spitzer Space Telescopes} as well as ground-based surveys have resulted in a boon in the numbers of high-redshift galaxies ($z>6$) identified by the Lyman-break in their optical and near-infrared colors. 
The high-redshift frontier is now firmly at $z\sim9-10$, the limit of \emph{HST} Lyman dropout technique, with a dozen high-fidelity candidates known \citep{Zheng12, Coe13, Bouwens11a,Bouwens11b, Bouwens13, Ellis13, Oesch13a, Oesch14}. These highest redshift candidates can be identified by their extremely red near-infrared colors ($J-H > 0.5$), a lack of flux in bluer (optical) bands, and --when available-- relatively blue H-4.5$\mu$m colors. The fainter $z\sim9-10$ candidates were found both behind lensing clusters \citep{Coe13,Zheng12}, and in ultra-deep WFC3/IR observations \citep{Bouwens11a,Ellis13, Oesch13a}.

The brightest objects at these redshifts are exceedingly rare but a number of them have been identified in both the CANDELS deep fields \citep{Oesch14} and more recently in \emph{HST} Pure-Parallel observations \citep{Calvi16,Morishita18a}. 
The existence of bright ($>L^*$) galaxies poses a critical challenge for early galaxy evolution during the first 500 Myr \citep[$10-30\times$ growth from $z\sim10$ to $z\sim7$,][]{Madau14, Bouwens15b, Ishigaki18}, as these few objects are too bright for the observed evolution of the luminosity function, hinting at a different formation mechanism of the brightest galaxies.

Their relative luminosity and spatial paucity should make these ideal targets for deep near  infrared imaging surveys conducted from the ground \citep[e.g. UltraVISTA][]{McCracken12}. However, their reliable detection and selection has proved challenging: 
20\% of the luminous $z\sim7$ candidate objects were rejected as spurious with HST follow-up \citep{Bowler15} and \cite{Stefanon17} could not reliably confirm any $z>8$ candidates using HST follow-up. 
Therefore, as the widest tier of the near-infrared imaging searches for $z>7$ bright sources, the BoRG survey stands the best chance to identify candidate $z\sim9-10$ galaxies.

There are three, mostly independent tests for the high-redshift nature of $z\sim9-10$ galaxy candidates: (a) one can obtain the Spitzer flux and color of these candidates to confirm their photometric redshift (\citealt{Roberts-Borsani16,Bridge19}, (b) one can observe the Lyman-$\alpha$ emission line if a (re)ionized bubble is present \citep{Zitrin15,Oesch15,Stark17,Larson18}, and (c) comparing these sizes against expectations for luminous galaxy candidates at $z\sim9-10$ and the sizes of potential interlopers \cite{Grazian12,Holwerda15}. 

The analytical models from \cite{Fall80} and \cite{Mo98} predict effective radii should scale with redshift somewhere between $\propto(1+z)^{-1}$ for galaxies living in halos of fixed mass or $\propto(1+z)^{-1.5}$ at a fixed circular velocity. Observational evidence from earlier samples also points to such scaling relations, with some studies preferring $(1 + z)^{-1}$ \citep{Bouwens04, Bouwens06,Oesch10a}, some studies preferring $(1+z)^{-1.5}$ \citep{Ferguson04}, and some
studies lying somewhere in between \citep{Hathi08,Ono13,Shibuya15}. For the bright ($>0.3L^*$) sources, one expects galaxies to follow $\propto(1+z)^{-1.5}$ relation \citep[cf][]{Holwerda15}.

In this paper, we examine the sizes of a sample of $z\sim9-10$ candidate galaxies identified in \cite{Morishita18a}. 
Our aim is to evaluate how well candidate galaxy size can be used in pre-selection of sources without relying on visual assessment. For near future larger area searches, the addition of an additional criterion other than color, will be invaluable.
This paper is organized as follows:
\S \ref{s:borg} describes the data and catalog used, 
\S \ref{s:color} explores the relation between \emph{HST} colors and effective radii for selected and rejected candidate $z\sim9-10$ galaxies, 
\S \ref{s:rez} shows the redshift evolution of the effective radii, 
\S \ref{s:SF} compares the inferred star-formation surface densities to previously identified bright sources, 
\S \ref{s:concl} briefly lists our concluding remarks.
We assume a flat cosmology of $H_0 = 73$ km/s/Mpc, $\Omega_0=0.28$. For $L^*$ luminosity, we use the z=3 value: $M_{1600}(z = 3) = 21.07$ \citep{Steidel99}.

\section{\label{s:borg}BoRG[z9] Survey Data}

Legacy field investigations with \emph{HST} have covered substantial area \citep[$>800$ arcmin$^2$][]{Grogin11,Koekemoer11}, and the latest $z>7$ samples are approaching $\sim1$k sources \citep[e.g.,][]{Bouwens15b,Finkelstein15}. However, at the bright end of the galaxy luminosities, the rarity of high-redshift candidate galaxies is problematic with their cosmic variance being the main source of uncertainty \citep{Barone-Nugent14}. A few large, contiguous fields can be significantly affected by cosmic variance \citep{Trenti08}.

The Brightest of the Reionizing Galaxies \emph{HST} survey \citep[BoRG][]{Trenti11,Trenti12, Bradley12, Schmidt14,Calvi16} has been designed specifically to contribute towards an unbiased measurement of the number density of the brightest galaxies at $z\sim8$ initially and now focuses on $z\sim9-10$, using \emph{HST} pure-parallel  opportunities to cover a comparable area with medium-deep optical and IR imaging but with effectively random pointings ($5\sigma$, $m_{AB}< 26.5$) over $\sim140$ independent sight lines so far.

\begin{figure*}
\centering
\includegraphics[height=0.33\textwidth]{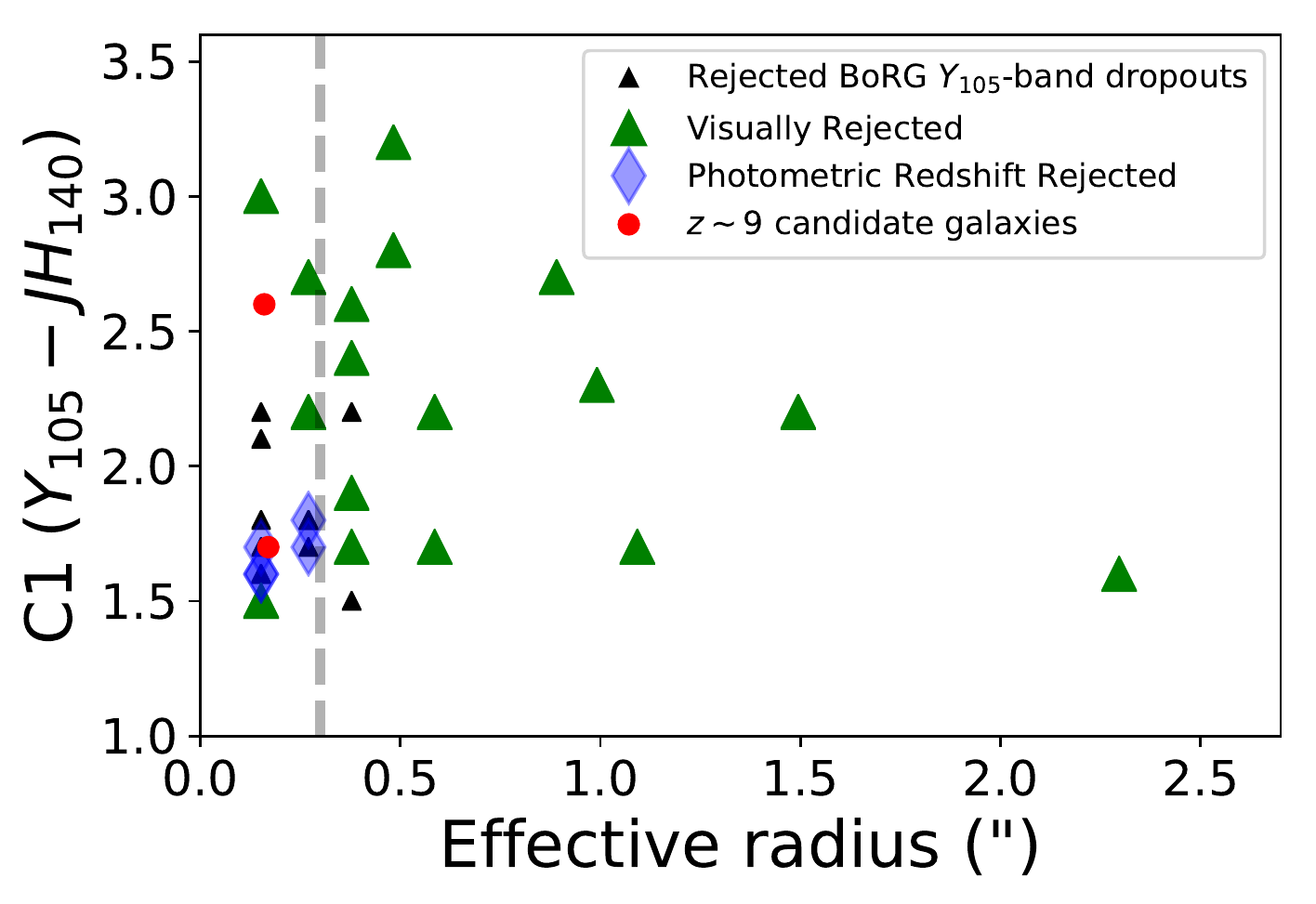}
\includegraphics[height=0.33\textwidth]{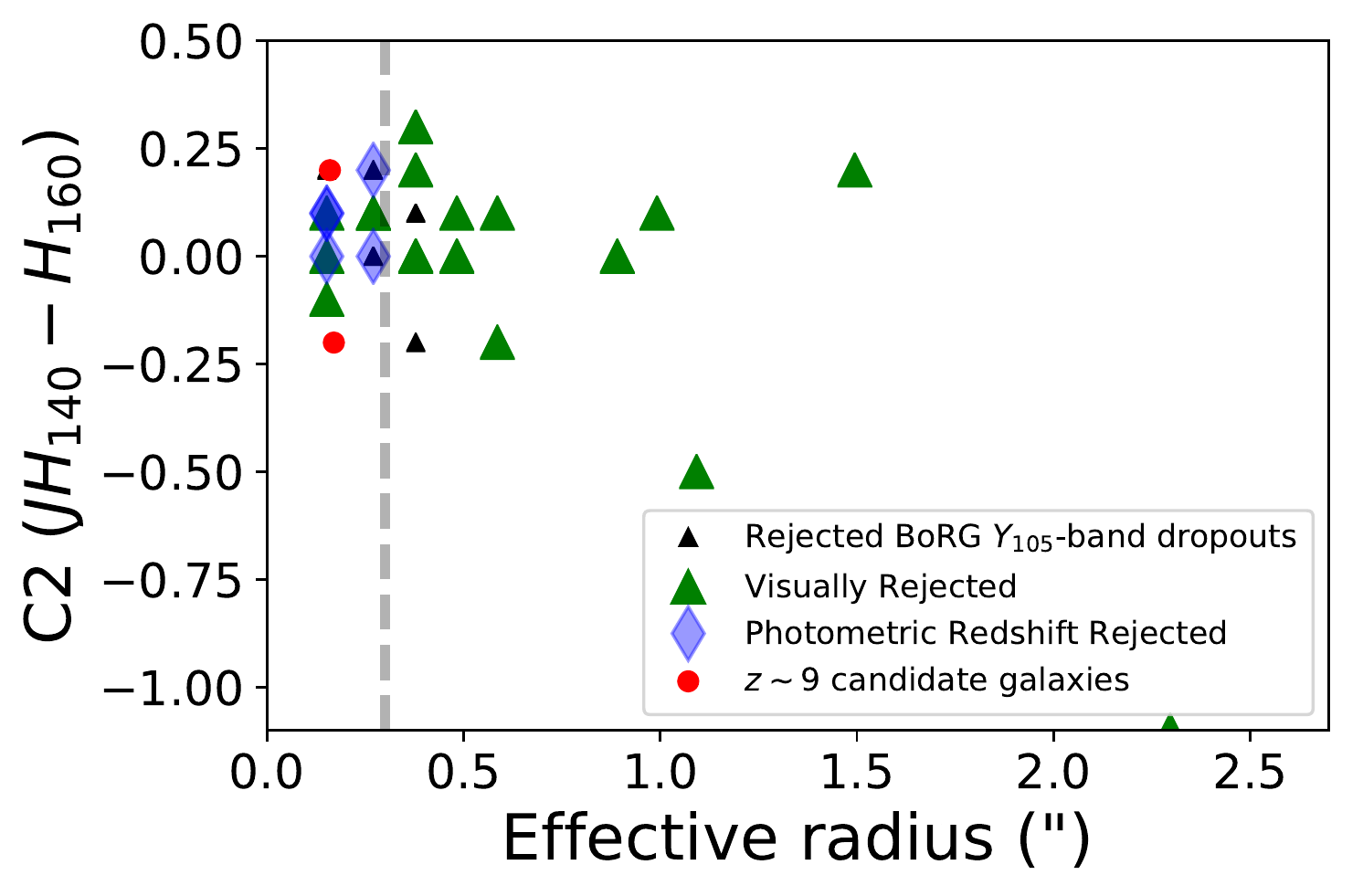}
\caption{The effective radii vs the two colors $Y_{105}-JH_{140}$ and $JH_{140}-H_{160}$ for the $Y_{105}$-band dropouts ($z\sim9$ candidates). Green triangles and blue diamonds are candidates rejected visually and/or through their photo-z solution. Black triangles were rejected for a different reason, mostly their H-3.6$\mu$m color. The red points are the selected $z\sim9$ candidate galaxies.}
\label{f:re_col:z9}
\end{figure*}
The initial BoRG survey aimed at the bright-end of the Luminosity Function (
LF) at $z\sim8$, using four filters on WFC3 \citep{Trenti11,Trenti12a, Bradley12, Bernard16,Schmidt14} explored 350 arcmin$^2$ and found 38 Y-band dropout candidates with $>L^*$, providing one of the strongest constraints on the $z\sim8$ LF shape. The next iteration of the survey, BoRG[z9] (GO 13767, PI. M. Trenti), which we use in this study, is optimized for higher redshift ($z>9$) galaxies with an updated set of 5 WFC3IR/UVIS filters (F300LP, F105W, F125W, F140W and F160W). Preliminary results from this survey were presented in \cite{Calvi16} and \cite{Morishita18a}. 
We use the latest vetted sample of $z\sim9$ and $z\sim10$ from 
\cite{Morishita18a}. They select $z\sim9$ galaxies from F105W (Y-band) dropouts and $z\sim10$ from F125W (J-band) dropouts. They combine the F140W (JH) and F160W (H-band) filter images to generate a detection image in which the size is determined by Source Extractor \citep{SE,seman}. 

\begin{figure*}
\centering
\includegraphics[height=0.34\textwidth]{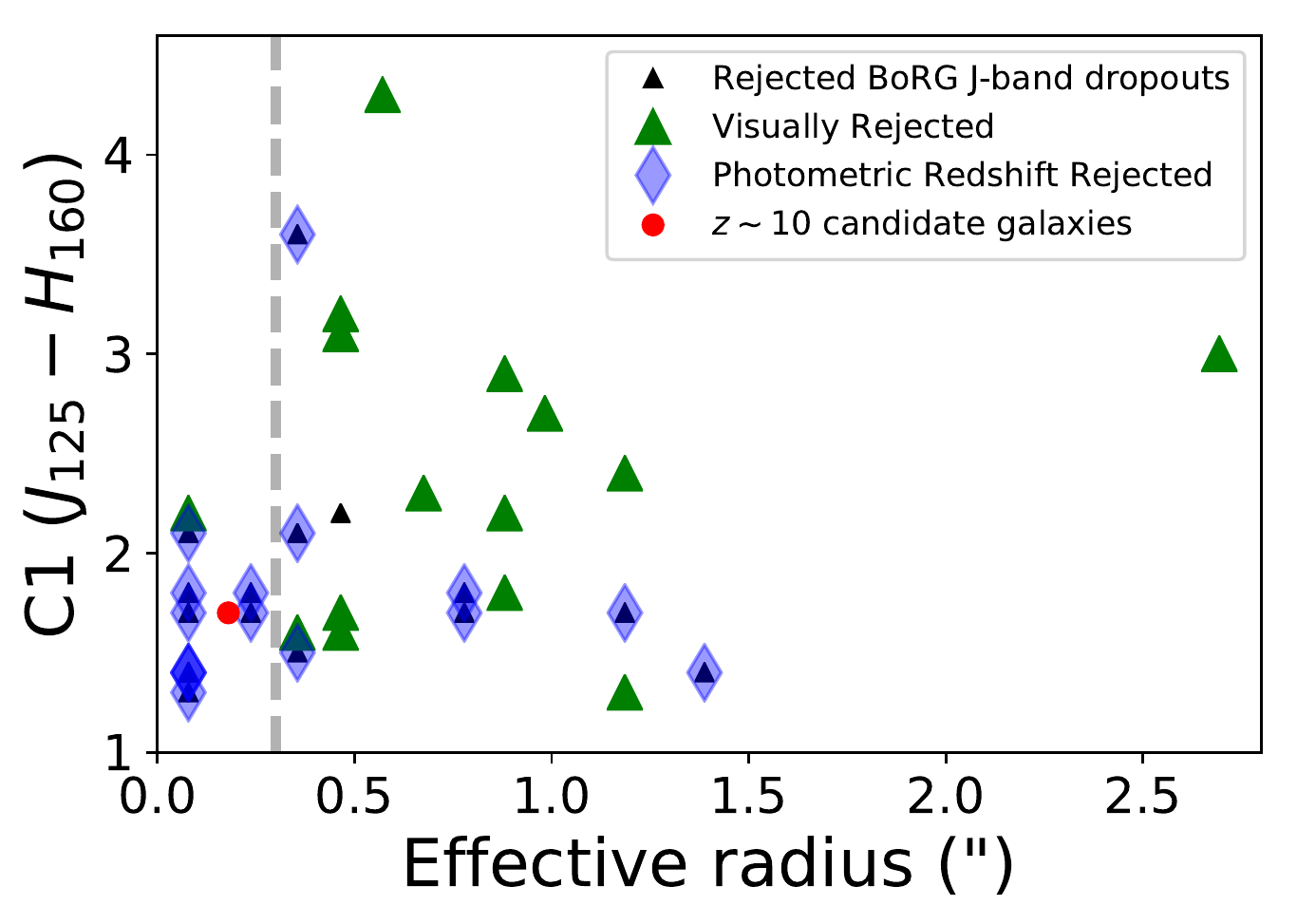}
\includegraphics[height=0.34\textwidth]{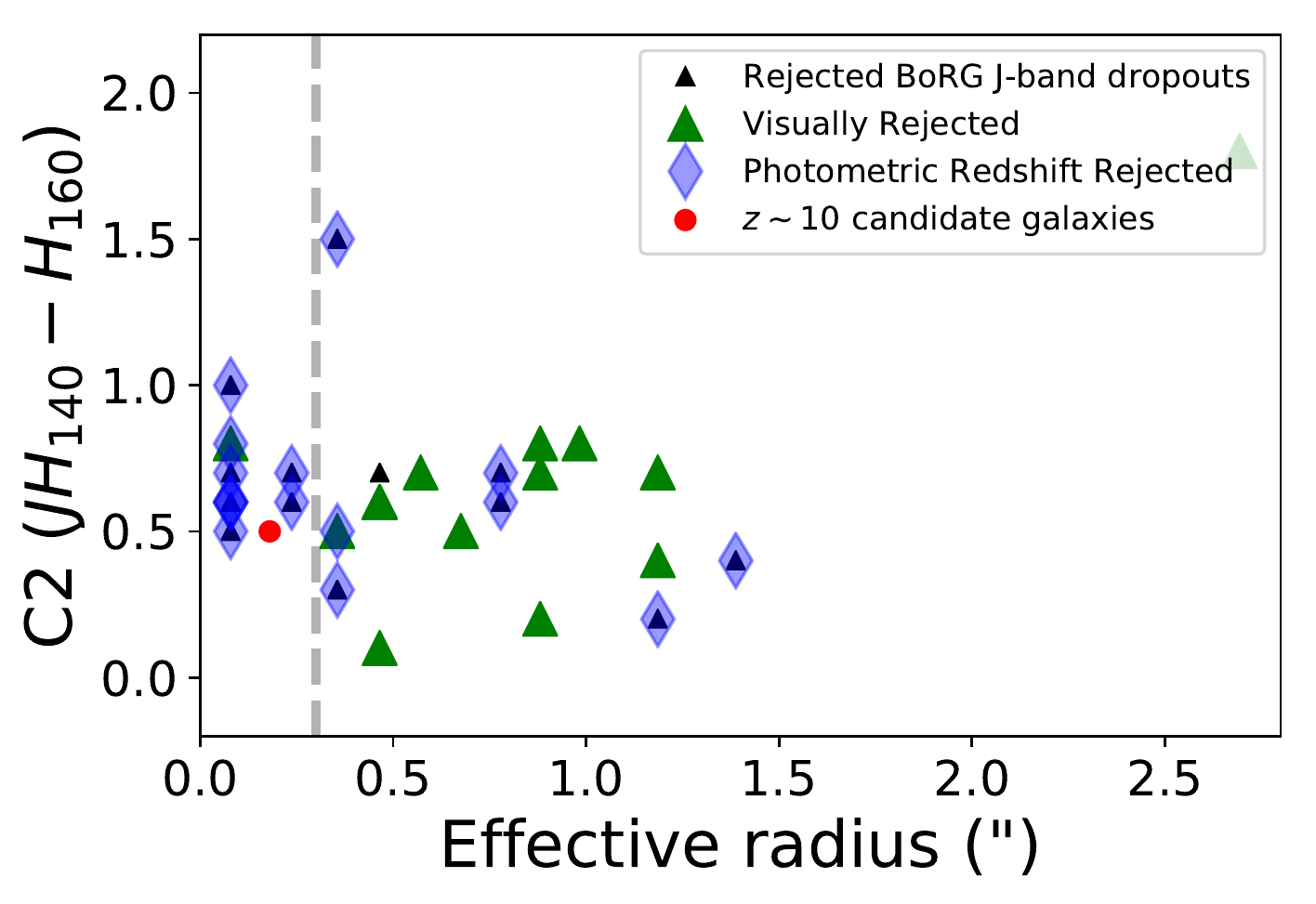}
\caption{The effective radii vs the two colors $J_{125}-H_{160}$ and $JH_{140}-H_{160}$ for the $J_{125}$-band dropouts ($z\sim10$ candidates). Green and blue triangles are candidates rejected visually or through their photo-z solution. Black triangles were rejected for a different reason, mostly their H-3.6$\mu$m. The red points are the selected $z\sim10$ candidate galaxies.}
\label{f:re_col:z10}
\end{figure*}
Source Extractor measures the half-light radii of objects through a simple growth curve approach: the pixels in an object stack are sorted, each pixel is 
reproduced ten times and the point within which 50\% of the flux is contained. This area (number of pixels) is then converted to a radius in arcseconds by $r_e = platescale \times \sqrt[]{A/2\pi}$.

Effective radii of high-redshift sources are typically determined with either {\sc galfit} \citep{galfit2} or source extractor \citep{SE,seman}. For drizzled images, like the CANDELS fields, {\sc Galfit} is the best option as it allows for different S\'ersic profiles and a separate sky-subtraction. However, in the case of undithered Hubble imaging, the sampling of the image is not acceptable for the galfit hard limits. Because the {\sc galfit} run parameters have to be fixed for small objects such as the ones studied here (e.g. fixed S\'ersic parameter and axis ratio) and the proven robustness of source extractor effective radii, we use these here. 

\cite{Huang13} compare the performance of source extractor and {\sc galfit} on artificial images at lower redshift and they find discrepancies in the effective radii of source extractor for the lower-luminosity but more extended objects ($m=26$, $r_e = $ 0\farcs48-0\farcs75) or brighter and more compact ($m=24$, $r_e < 0\farcs05$) than those presented here. In the former case, source extractor over-estimates effective radius and the in latter case, under-estimates it. We caution that --once there is enough information for a profile fit-- {\sc galfit} will be less biased than source extractor but for the luminosity and size we are working with here, the difference will be minimal \citep[see e.g.][]{Oesch10a}. 
We convert these effective radii to kpc by correcting for the PSF: $r_e (cor) = \sqrt[]{r_e^2 - 0.13^2}$ (the WFC3 PSF is 0\farcs13 for $H_{160}$) and converting it to kpc for the appropriate photometric redshift as determined in the $H_{160}$ combined image by \cite{Morishita18a}. \cite{Morishita18a} corrected their effective radii and luminosities already for the mild but significant lensing magnification they found in these galaxies by following the prescription by \cite{Mason15b}.

\section{\label{s:color}\emph{HST} Colors and Sizes}

\cite{Morishita18a} vetted their dropout galaxies through a combination of photo-z fits, visual inspection (including a size criterion), and complementary Spitzer observations when available. We use the complete list of candidate galaxies without the visual criterion of \cite{Morishita18a} applied.

A first check is to see if the \emph{HST} colors and the effective radii of these candidate galaxies separate out the selected sources and the rejected sources. \cite{Morishita18a} present a color C1 and C2 in their tables. C1 is the $J_{125}-H_{160}$ or the $Y_{105}-JH_{140}$ for J- and Y-band dropouts, respectively $z\sim9$ and $z\sim10$ candidate galaxies. C2 is $JH_{140}-H_{160}$ for all objects.

Figure \ref{f:re_col:z9} shows colors and on-sky effective radii of the candidate sources and those selected by \cite{Morishita18a} as very probably $z\sim9$ sources. Selection of the $z\sim9$ galaxies is problematic to reproduce with a color-$r_e$ selection based on \emph{HST} colors. In \cite{Holwerda15}, it was shown that the $z\sim9$ selection could be vetted with either effective radius or the H-[3.6] color and it was argued that effective radius could perhaps stand in for this color if not available. Figure \ref{f:re_col:z9} illustrates the problem with a single value size cut as a hard size criterion would effectively cull some interlopers but it would still leave a majority of contaminants. However, a declining probability of inclusion with size --which is implicitly done a visual inspection-- could supplant the visual inspection in supplementing the photo-z rejections. 

Figure \ref{f:re_col:z10} shows the two colors and the effective radii for the candidate objects and the single vetted $z\sim10$ object. Here, the BoRG filter set does suggest a color-effective radius as a possible way to select highly likely $z\sim10$ candidate galaxies. J-band dropouts are selected using a $J_{125}-H_{160} > 1.3$ and s/n constraints. In Figure \ref{f:re_col:z10} one could require $r_e <0\farcs3$ and $JH_{140} - H_{160} < 0.55$ would cull most of interlopers that now were rejected visually, through their photometric redshift, or IRAC color. We argue here that the size may act as a prior to be included in photometric redshift selection. 
The strict effective radius criterion ($r_e <0\farcs3$, where $r_e$ is corrected for the PSF), results in a removal of 65-70\% of the visual rejections and $\sim50$\% of the all rejections in the z=9--10 samples.

A size criterion does select against the brightest galaxies in a luminosity function due to the size-luminosity relation. At $z\sim10$, an explicit criterion --rather than a more implicit visual one-- of $0\farcs3$ translates to 1.25 kpc limit and if earlier size-luminosity relations hold \citep[e.g.,][at $z\sim7$]{Grazian12} an effective cutoff at $M^* = 10^{10} M_\odot$ or $M_{UV} \sim -22.9$ or alternatively $M_{UV} \sim -28.2$ assuming the \cite{Holwerda15} size-luminosity relation.

We argue that it is better to have this explicit bias with a cut in effective radius rather than an implicit size selection based on visual inspection. The absolute luminosity selected against however is predicted to be exceedingly rare at $z\sim9-10$ \citep[see][]{Mason15,Bouwens15b,Finkelstein15}. 

One can compute the relative distribution of sizes of galaxies based on the simulated luminosity function of \cite{Mason15}, ranging from 3 mag brighter than $M^*$, M=-24.5 to M=-10.5, with an assumed size-luminosity relation, either the one from \cite{Grazian12}, for $z=7$ galaxies or the one from \cite{Holwerda15} for bright $z=9-10$ galaxies. There are uncertainties in the luminosity function parameters from \cite{Mason15} as well as in the assumed luminosity-size relation. In addition, there is a measurement error in the effective radius which one can approximate with the WFC3 PSF width. Bootstrapping these into an uncertainty for the size function is illustrated in Appendix A. 

Depending on the assumed size-luminosity relation, a size criterion for $z>7$ galaxies of $0\farcs3$ ($\sim$1.25 kpc at $z=9$) removes a  small fraction of the total expected $z>7$ population  (see Appendix and Figure \ref{f:SizeFunction}). Figure \ref{f:SizeFunction2} shows the total number of galaxies in the high redshift Universe as a function of apparent size for z=7 and z=9 assuming either the \cite{Holwerda15} or the \cite{Grazian12} size-luminosity relation to show the spread in the size function. 
These examples illustrate the role the assumed size-luminosity relation plays in the loss estimate of an explicit size cut. 

Given how uncertain the size-luminosity relation is at $z=9-10$, any implicit (visual inspection) or explicit size criterion should be accounted for. Typical photometric contamination of high-redshift candidate galaxies is still $\sim10-50$\% (see Figure \ref{f:re_col:z9} and \ref{f:re_col:z10}). For comparison, the fraction of galaxies removed in a survey by a hard size cut depends strongly on the size-luminosity relation assumed and how far one is willing to extrapolate the luminosity function by \cite{Mason15}. If this is taken \textit{in extremis}, one extends the luminosity function so far there is no longer enough volume in the early Universe to support a single galaxy of that size according to the luminosity function (illustrated in Fig. \ref{f:SizeFunction2}). 

\begin{figure*}
    \centering
    \includegraphics[width=0.49\textwidth]{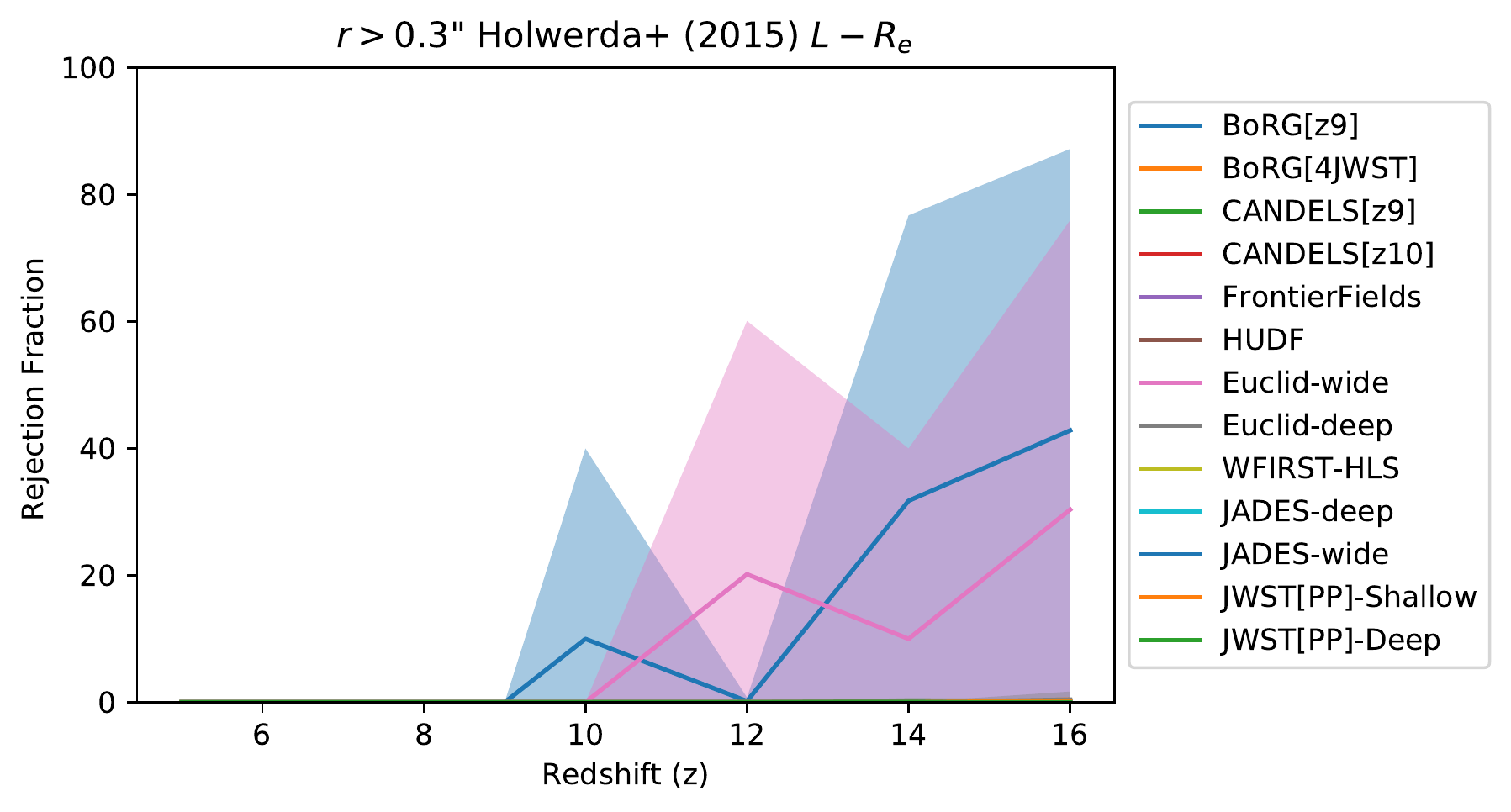}
    \includegraphics[width=0.49\textwidth]{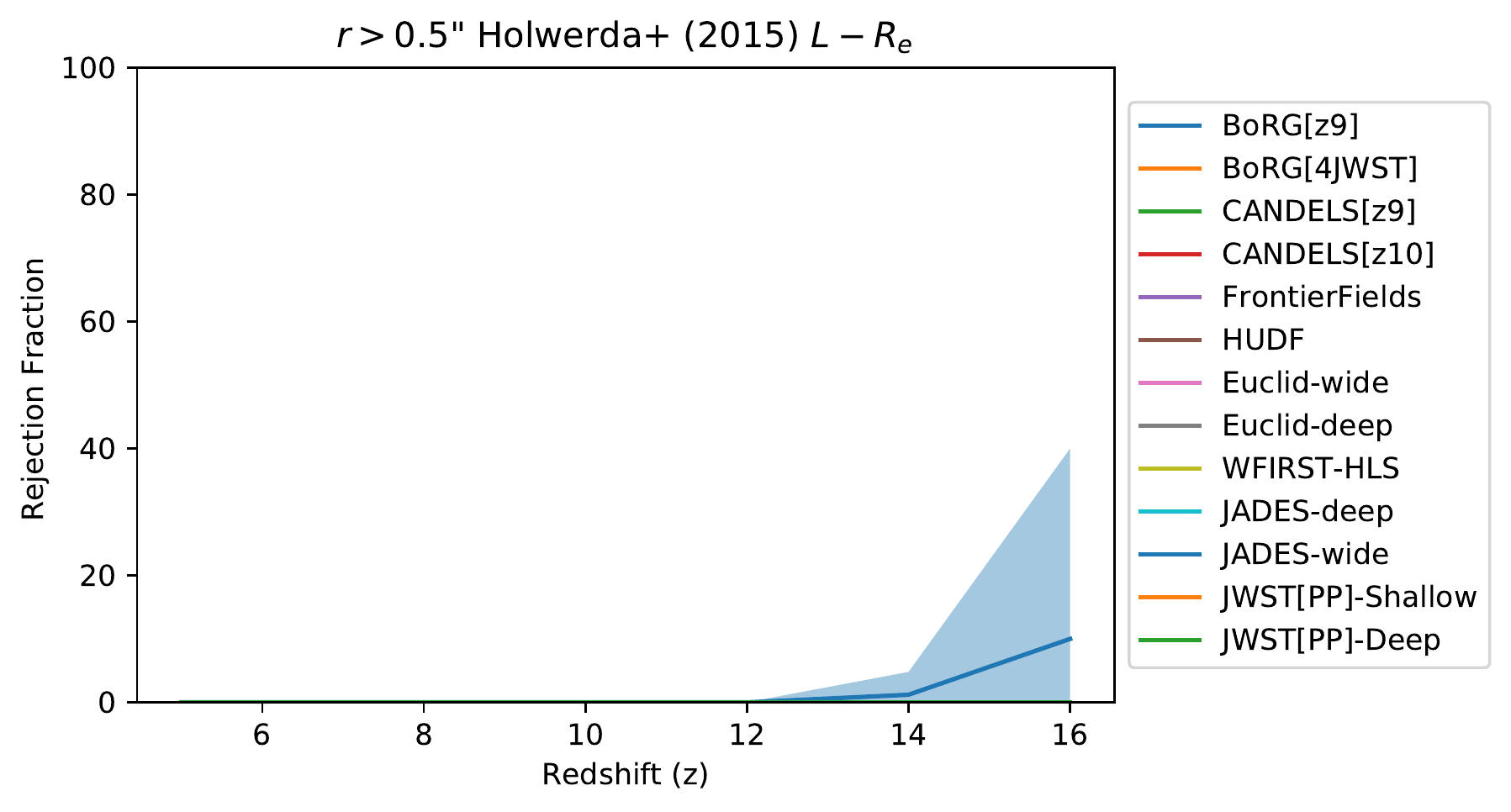}
    \caption{The rejection percentage of different surveys, assuming the \cite{Holwerda15} luminosity-size relation and either a $r_e = 0\farcs3$ or $0\farcs5$ (right) selection criteria. BoRG[z9] rejection fractions reach 10\% at z=10 for 0.3".}
    \label{f:rejectfrac}
\end{figure*}

If one assumes just the luminosity range calculated in \cite{Mason15} as a viable range, the rejection rate is less than 1ppm for either luminosity-size relation. 
If one limits oneself to 6 magnitudes brighter than the Mason et al. range and limit to the BoRG[z9] detection limit \citep[$m_H = 23.6$][]{Rojas-Ruiz20}, one gets the rejection fraction of the total galaxy population tabulated in Table \ref{t:lossrate}. Assuming the \cite{Holwerda15} luminosity-size relation for bright sources at z=9, the rejection rate by either size cut above z=5 is much smaller compared to the contamination rate by interlopers from photometry alone within expected scatter (Table \ref{t:lossrate} and Figure \ref{f:rejectfrac}). However, the loss rate is similar or worse to the contamination rate if one assumes the \cite{Grazian12} luminosity-size relation for z=7. 

For candidates found in BoRG[z9], the luminosity-size relation for bright galaxies from \cite{Holwerda15} is the most appropriate for z=9 candidate objects. If one uses the 0\farcs3 cut on BoRG[z9], this results in a mean rejection rate of 10\% at z=10 (Figure \ref{f:rejectfrac}, Table \ref{t:lossrate}). For comparison, close to half the candidate sources in \cite{Morishita18a} was rejected visually (Figures \ref{f:re_col:z9} and \ref{f:re_col:z10}) making contamination from lower redshift sources a much greater concern.

We explored all other relevant discovery surveys (Euclid, WFIRST, JWST and HST CANDELS) to explore if other surveys would suffer similar rejection rates of bona-fide high-z sources. Only Euclid would suffer from this with the strictest 0\farcs3 size cut \citep[assuming the luminosity-size relation from ][]{Holwerda15}, and only above z=10 (Figure \ref{f:rejectfrac}).

{}

Moving beyond hard cuts for high-redshift Lymann-break object selection, we would recommend a probabilistic approach with combined probability function over color, size, and perhaps including morphometrics \citep[e.g.][using asymmetry and Gini]{Kusmic19} for candidate high redshift source selection from source catalogs. In the lower redshift regime, photometric redshift estimates have included morphological information in addition to colors to improve reliability \citep[ e.g.][]{Xia09,Wilson19a,Paul18,Soo17,Momcheva16a}.

Possible applications include GO$^2$LF, the Great Observatories Square-degree Legacy Fields \citep{Stefanon19b} or future searches using, HST, JWST, EUCLID or WFIRST imaging, to include both morphometrics (Gini and asymmetry) as well as effective radius (e.g. a sharply declining prior above 0\farcs3 and a 0\farcs5 cut-off) for pre-selection of high redshift candidate galaxies to substitute or prescreen before a visual inspection.

\begin{figure}
\centering
\includegraphics[width=0.47\textwidth]{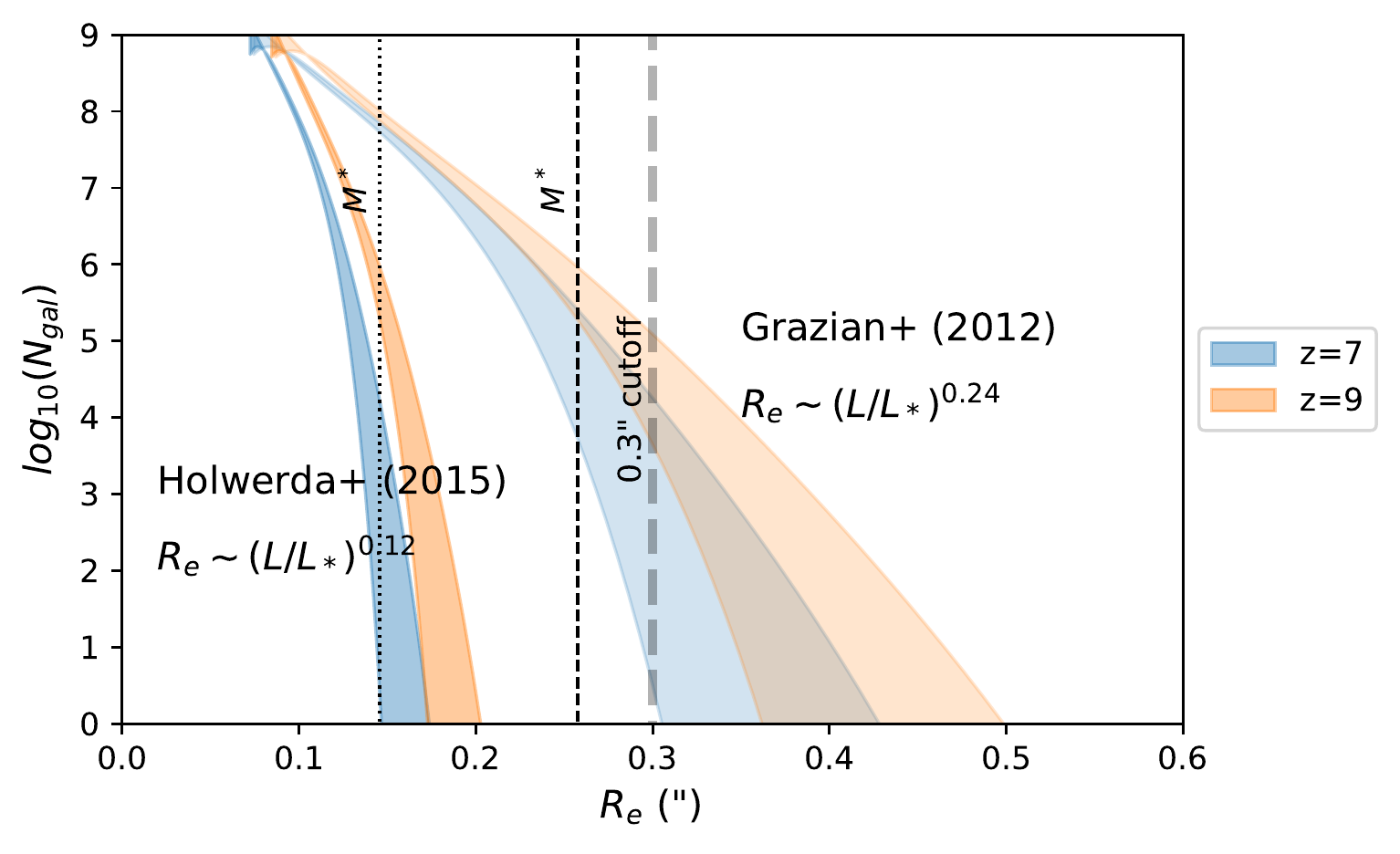}
\caption{The total number of galaxies in the early Universe as a function of observed effective radius, assuming the luminosity functions from \cite{Mason15} evaluated from M=-38.5 ($\sim14$ mag brighter than $M^*$) to M=-10.5, the two size-luminosity functions at higher redshift, (\cite{Holwerda15} and \cite{Grazian12}), and the volume available at each redshift. Width of the relation is an assumed error the width of the WFC3 PSF (0\farcs1). The narrow lines are the point of $M^*$ of the \cite{Mason15} luminosity function (dotted and dashed for the relative size-luminosity relations respectively). The size cut of 0\farcs3 would remove a very small fraction ($<<$1ppm) of galaxies at $z\sim9-10$ assuming the $z\sim7$ size-luminosity relation from \cite{Grazian12}, a 0\farcs5 hard size cut would remove none.  }
\label{f:SizeFunction2}
\end{figure}

\section{\label{s:rez}Effective radius with redshift}
\begin{figure*}
\centering
\includegraphics[width=\textwidth]{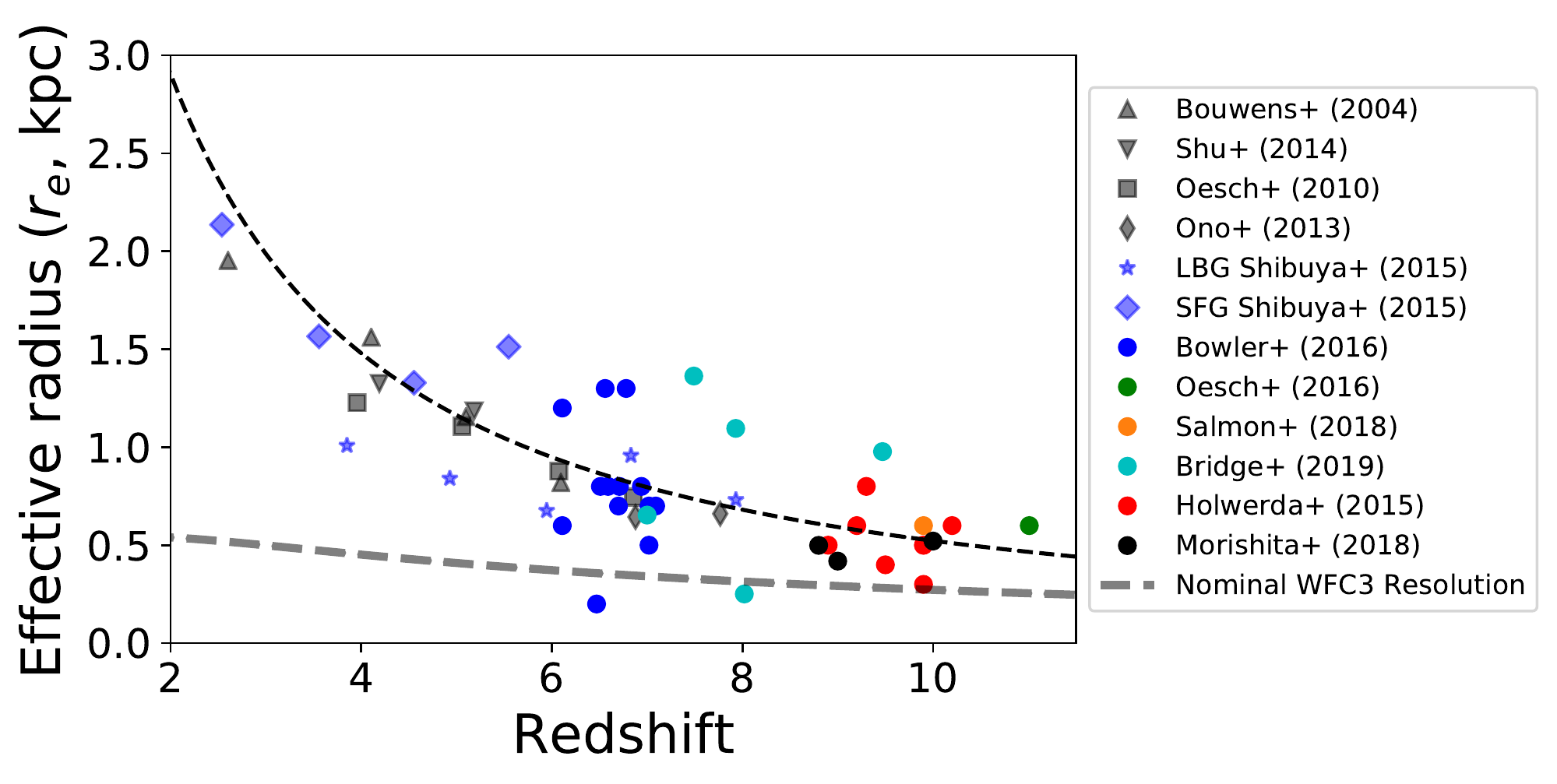}
\caption{The relation between effective radius and redshift for the bright galaxies ($L>L^*_{z=3}$) from several previous studies: mean effective radii from
 \cite{Bouwens04}, \cite{Zheng14c}, \cite{Oesch10a}, \cite{Ono13} (gray symbols). The mean values of the size distributions for the star-forming galaxies (SFG) and Lyman Break Galaxies (LBG) from \cite{Shibuya15} for comparison (light blue symbols). The effective radii as a function of redshift for individual galaxies from \cite{Holwerda15} ($z\sim9$), \cite{Bowler17} and \cite{Bridge19} ($z\sim8$), and the $z\sim9-10$ galaxies identified by \cite{Morishita18a} that we discuss in the paper (circles). The dashed line is the best fit to $r_{e} \propto (1=z)^m$ from \cite{Holwerda15} with $m = -1.32$, a value somewhere in between the two extreme cases. The mean values are for bright ($L>0.3L^*$) galaxies while the mode is measured over the whole observed luminosity range but is dominated by lower luminosity ($L<0.3L^*$) galaxies. The individual objects at the higher redshifts are predominantly bright and show a wide spread around the mean for bright sources.}
\label{f:rez}
\end{figure*}
\cite{Morishita18a} computed the effective radius from the F160W ($H_{160}$) image, similar to most other studies on the size of high-redshift which use $H_{160}$ \citep[e.g.][]{Shibuya15,Holwerda15}. 
We therefore expect the source extractor sizes to align reasonably well with the {\sc galfit} fit sizes \citep{galfit,galfit2} of previous studies.  \cite{Morishita18a} corrected their luminosities and sizes already for the lensing magnification they determined affected the majority of their candidate $z\sim9-10$ galaxies following the prescription in \cite{Mason15}.

Figure \ref{f:rez} compares the effective radii from \cite{Morishita18a} to individual candidate galaxy sizes found by \cite{Holwerda15}, \cite{Bowler17}, \cite{Oesch16}, \cite{Salmon18}, and \cite{Bridge19}, the mean values from \cite{Bouwens04}, \cite{Zheng14c}, \cite{Oesch10a}, \cite{Ono13} and the mode values from \cite{Shibuya15} for both star-forming galaxies (SFG) and Lyman Break Galaxies (LBG). The best fit assuming $r_{e} \propto (1=z)^m$ from \cite{Holwerda15} with $m = -1.32$ is also shown (thin dashed line), as is the nominal resolution of WFC3 (wide dashed gray line).

\begin{figure}
\centering
\includegraphics[width=0.5\textwidth]{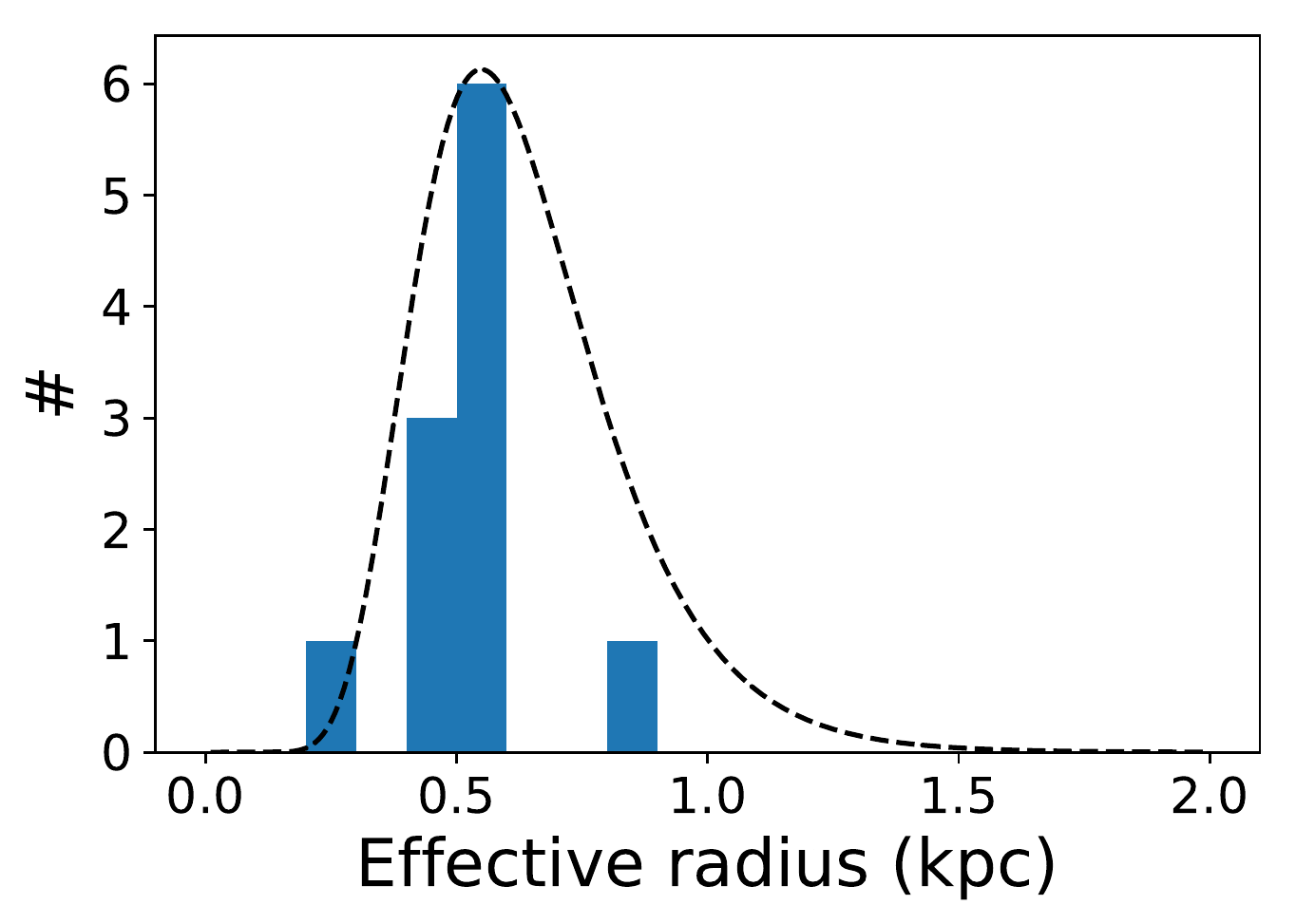}
\caption{The histogram of the sizes of $z\sim9-10$ galaxies as identified by \cite{Holwerda15} and \cite{Morishita18a} combined. The histogram follows a peak and tail to higher values of a lognormal distribution that \cite{Shibuya15} points out works well at lower redshifts ($z<6$) for the size distribution of galaxies.}
\label{f:rehist}
\end{figure}
\cite{Shibuya15} already pointed out that the distribution of galaxy sizes is not well characterized by a mean value but more accurately by a peak, characterized by the mode and a tail to higher values. 
Figure \ref{f:rehist} shows the histogram of the effective radii of $z\sim9-10$ candidate galaxies from \cite{Holwerda15} and \cite{Morishita18a}. 
The combined data-set hints at the same peak and possibly a tail to higher values similar to the one \cite{Shibuya15} identify at $z\sim6$ redshifts where the statistics are better. The $z\sim9$ show similar peak size as $z\sim6$ but a lack of an extended wing thus far. 

Low number statistics remain an issue and the current distribution of bright $z\sim9$ galaxies is consistent with the distribution at $z\sim6$. \cite{Curtis-Lake16} point to a lack of size evolution $z>4$ but as \cite{Shibuya15} noted, this is more the effect of the choice between mean or mode used in characterizing the size evolution. The mode of the size distribution evolves slowly with redshift and most of the more rapid evolution occurring in the tail of the size distribution, influencing the mean.
\begin{figure*}
\centering
\includegraphics[width=\textwidth]{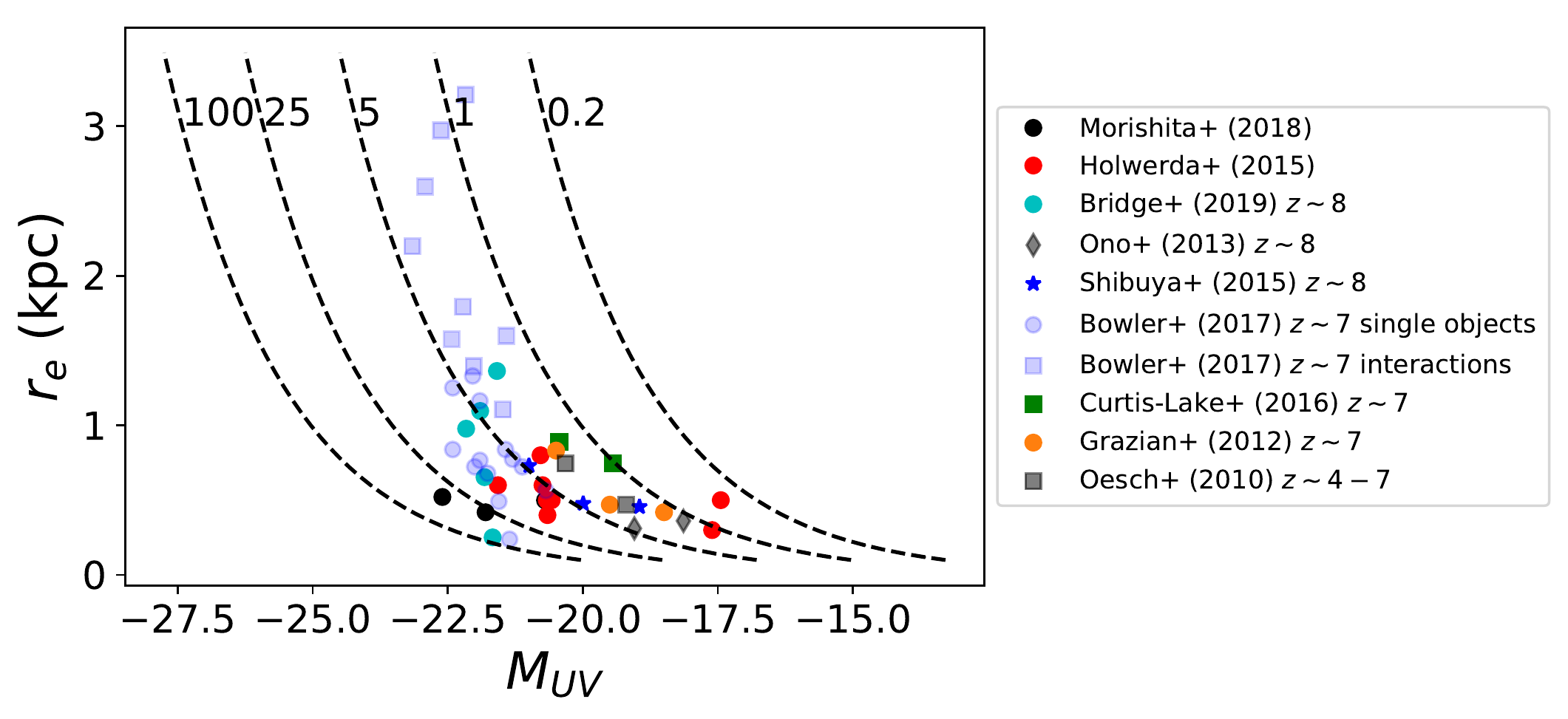}
\caption{The absolute restframe ultraviolet magnitude versus effective radius in kiloparsec for \cite{Holwerda15} and \cite{Morishita18a} with lines of constant star-formation surface densities ($\rm \Sigma_{SFR} (M_\odot/kpc^2)$). The candidate galaxies from \cite{Morishita18a} appear to be at similar star-formation densities than the sample found in \cite{Holwerda15}. For comparison, the $z\sim7$ $r_e - M_{UV}$ relations from \cite{Bridge19}, \cite{Ono13}, \cite{Bowler17}, \cite{Curtis-Lake16}, \cite{Grazian17} and \cite{Shibuya15} are shown as well. \cite{Bowler17} distinguishes between single objects and interactions. The \cite{Morishita18a} and \cite{Holwerda15} $z\sim9$ objects lie in the range for individual objects at $z\sim7$. 
}
\label{f:Muvre}
\end{figure*}

The gradual size evolution from $z\sim9$ to lower redshifts appear in contrast to the potential strong evolution in the luminosity function found by \citep{Bouwens15b,Oesch18a}. However, \cite{Morishita18a} already point out that their results on the brightest end of the luminosity function ($L>0.3L^*$) are completely consistent with both other BoRG results \citep{Calvi16} and other searches for high redshift galaxies \citep{Oesch13,Bouwens15b, Bernard16, Oesch18a, Livermore18} as well as theoretical predictions  \cite[e.g.,][]{Mason15}. The majority of the strong evolution is happening at fainter luminosities than BoRG probes. 

\section{\label{s:SF}Star-formation Surface density}

Figure \ref{f:Muvre} shows the effective radius as a function of absolute magnitude in restframe ultraviolet with lines of constant star-formation surface density marked following the relation in \cite{Ono13}. Neither dust extinction nor strong emission lines are assumed in this simple relation between size, absolute UV luminosity and star-formation surface density.

The $z\sim9-10$ candidate galaxies from \cite{Holwerda15} and \cite{Morishita18a} are similar in star-formation surface density. Two exceptions are suggestive of much higher values ($\rm \Sigma_{SFR}>25 M_\odot / kpc^2$): 0956+2848-98 and 2229-0945-394.
These higher star-formation surface density candidate galaxies could be prime targets for spectroscopic follow-up with the aim to detect either the Lyman-$\alpha$ emission line or with the James Webb Space Telescope for nebular emission lines.

Figure \ref{f:Muvre} compares the star-formation surface densities found for the \cite{Holwerda15} and \cite{Morishita18a} $z\sim9$ objects to those found at lower redshift \citep[$z\sim7$][]{Bowler17,Curtis-Lake16,Grazian17}. The comparison to the \cite{Bowler17} is instructive to show that the $z\sim9$ sources are of comparable size and luminosity as the single $z\sim7$ objects. 

\cite{Kusmic19} find that the morphometrics (Asymmetry and Gini) of $z\sim8$ candidate galaxies presented in \cite{Bridge19} are consistent with mostly unperturbed galaxies in \cite{Lotz10a,Curtis-Lake16} but not with smooth S\'ersic profiles. Our size-luminosity findings are consistent with no major mergers in this population: only very gradual evolution is needed to transition from these $z\sim9$ sample of objects to the brighter, single $z\sim7-8$ sources reported in \cite{Bowler17} and  \cite{Bridge19}. 

We note here that higher redshift selection criteria, be it an explicit size criterion or a visual inspection may select against potential ongoing mergers. Our size measurements are consistent with single objects only. This is not to be taken as evidence of an absence of galaxy mergers, just consistent with no mergers being selected at z=9. 

\section{\label{s:concl}Concluding Remarks}

We conclude that no clear \emph{HST} color and effective radius cuts can really replace the visual inspection and photometric redshift check that \cite{Morishita18a} performed to vet all the initial candidates selected on drop-out color for $z\sim9$ ($Y_{105}$-dropouts) and $z\sim10$ ($J_{125}$-dropouts). For the $J_{125}$-dropouts, a color-effective radius cut may be possible but the numbers are insufficient for a well motivated limit (Figure \ref{f:re_col:z10}). 

The selection with effective radius, rather than subjective one in a visual inspection may remove some of the ambiguity in selection however. An explicit size cut may remove the very brightest objects ($L >> L^*$) in a survey, depending on the actual size-luminosity relation of the epoch. For example, the 0\farcs3 cut removes objects brighter than 2 mag than $M^*$ at z=7-10 (Figure \ref{f:SizeFunction}) but also 65-70\% of the visual rejections in the Morishita sample. 

Rather than hard cuts in both color and effective radius, a probabilistic approach for Lymann break selection can be now considered. The prior for effective radius can drop off to zero between 0\farcs3 and 0\farcs5 for near-future HST-quality infrared datasets searches for high redshift, Lymann break candidates such as GO$^2$LF, JWST, targeted and parallel fields, EUCLID and WFIRST imaging. 

The reported sizes from \cite{Morishita18a} are all consistent with z=9 and z=10 objects. The vetted sample from \cite{Morishita18a} appears to have successfully doubled the number of known bright candidate $z\sim9-10$ galaxies, ideal for future spectroscopic follow-up observations with e.g. JWST and Keck.

The effective radii of the \cite{Morishita18a} $z\sim9-10$ galaxies are following the previously observed trend of the mean effective radius of high-redshift bright galaxies, following $r_e\propto(1+z)^{-1.3}$ (Figure \ref{f:rez}). The distribution of all the $z\sim9-10$ galaxies also suggest strongly that the lognormal distribution of bright galaxy sizes observed by \cite{Shibuya15} holds out to this redshift (Figure \ref{f:rehist}).

The star-formation surface densities implied by the size and luminosities of these $z\sim9-10$ galaxies is very comparable to those found in previous work on bright galaxies at these redshifts (Figure \ref{f:Muvre}). In part this can be attributed to the selections by \cite{Morishita18a} against large galaxies. Size measurement of photometrically selected high-redshift galaxies remain a valuable and impartial \textit{a posteriori} check of their high-redshift nature.

\section*{Acknowledgements}

We thank the anonymous referee for their well thought-out, constructive, and thorough comments on earlier drafts. Their effort and contributions are much appreciated. 

This work was supported by a NASA Keck PI Data Award, administered by the NASA Exoplanet Science Institute. Data presented herein were obtained at the W. M. Keck Observatory from telescope time allocated to the National Aeronautics and Space Administration through the agency's scientific partnership with the California Institute of Technology and the University of California. The Observatory was made possible by the generous financial support of the W. M. Keck Foundation.

The authors wish to recognize and acknowledge the very significant cultural role and reverence that the summit of Mauna Kea has always had within the indigenous Hawaiian community. We are most fortunate to have the opportunity to conduct observations from this mountain.

This research has made use of the NASA/IPAC Extragalactic Database (NED) which is operated by the Jet Propulsion Laboratory, California Institute of Technology, under contract with the National Aeronautics and Space Administration. 
This research has made use of NASA's Astrophysics Data System.
This research made use of Astropy, a community-developed core Python package for Astronomy \citep{Astropy-Collaboration13a}. This research made use of matplotlib, a Python library for publication quality graphics \citep{Hunter07}. PyRAF is a product of the Space Telescope Science Institute, which is operated by AURA for NASA. This research made use of SciPy \citep{scipy}.


\appendix

\section{Simulating the loss rate from a hard size criteria}

\begin{figure}[h]
\centering
\includegraphics[width=\textwidth]{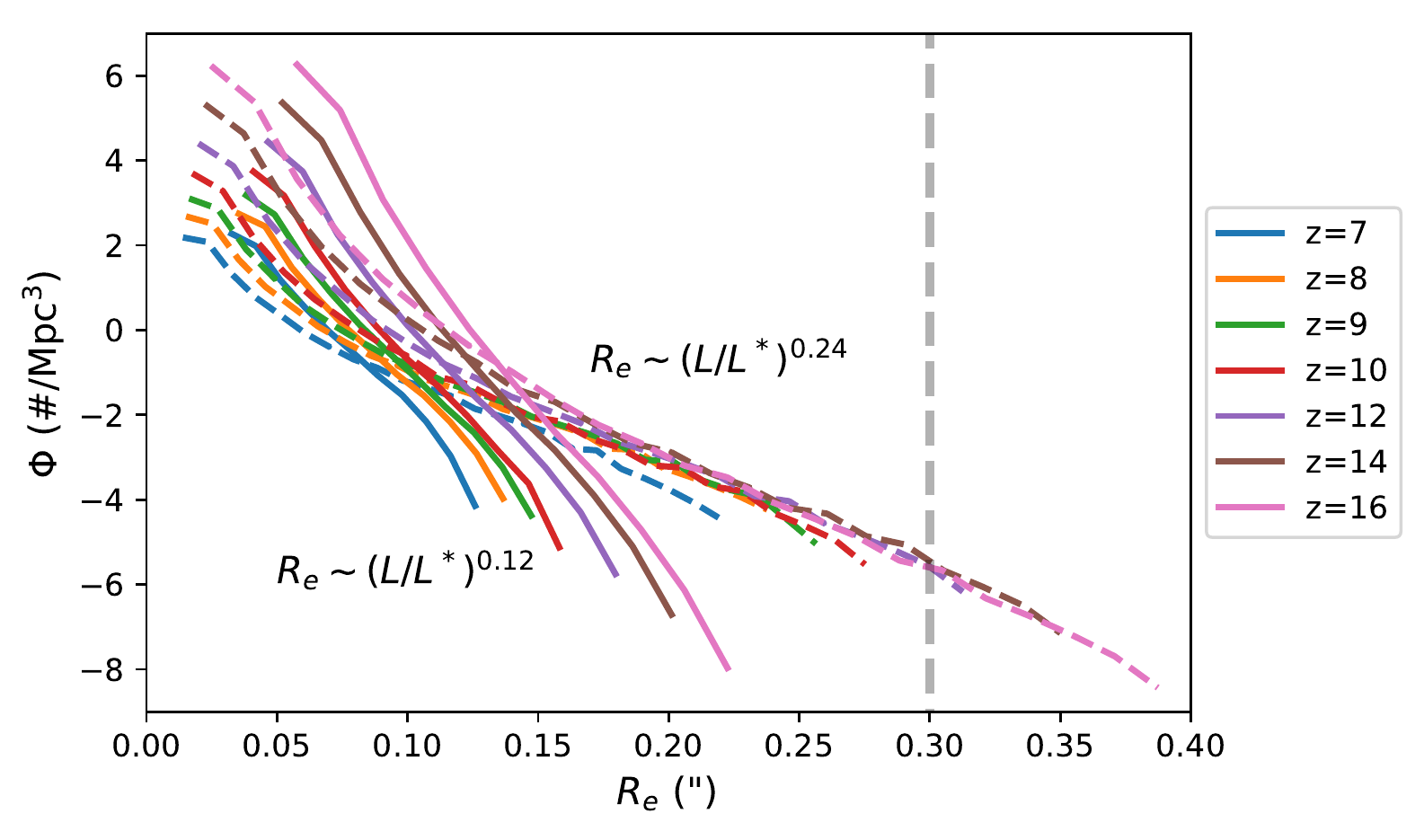}
\caption{The distribution of observed effective radii, assuming the luminosity functions from \cite{Mason15} evaluated over M=-22.5 ($\sim2$ mag brighter than $M^*$) to M=-10.5 and two size-luminosity relations found at high redshift. The solid lines use the size-luminosity relation $(R\sim L/L*)^{0.12}$, found for $z=9-10$ galaxies in \cite{Holwerda15} and the dashed lines the $z=7$ size-luminosity relation $(R\sim L/L*)^{0.24}$, presented in \cite{Grazian12}. \cite{Ono13} found a much flatter distribution of sizes for high redshift populations but thanks to the evolution in the luminosity function, $>$0\farcs3 galaxies would still be very rare (Table \ref{t:lossrate}).}
\label{f:SizeFunction}
\end{figure}

To estimate the loss rate from a hard size cut, we bootstrap the numbers of galaxies in the early Universe using the \cite{Mason15} luminosity function (their Table 1 and Figure 8). We evaluate two size-luminosity relations, the one from \cite{Grazian12} for z=7 galaxies and the one from \cite{Holwerda15} for z=9 population of galaxies to translate the evolution in the luminosity function into a size distribution at different redshifts.

The uncertainties in the luminosity function parameters ($\alpha$, $M^*$ and $log(\phi)$ are given in \cite{Mason15}, their Table 1, the first two symmetric and the last one asymmetric. The uncertainties in the luminosity-size relations (normalization and exponent) are listed in \cite{Holwerda15}, Table 3, all assumed to be symmetric. 
We bootstrap the size distributions at each redshift in the \cite{Mason15} luminosity functions 1000 times randomly varying the luminosity function and the luminosity-size relation according to their respective errors. The size distribution function is then subjected to a measurement error of 0\farcs1. This is a fiducial error as many packages give no error ({\sc source extractor}), or a underestimated error based on a $\chi^2$ value (e.g. {\sc galfit}). It is the FWHM of the NIR instrumentation on HST and we assume it to be representative of the error in future observatories (WFIRST, EUCLID and JWST/NIRCam) at the same wavelengths.

\begin{figure*}
\centering
\includegraphics[width=0.49\textwidth]{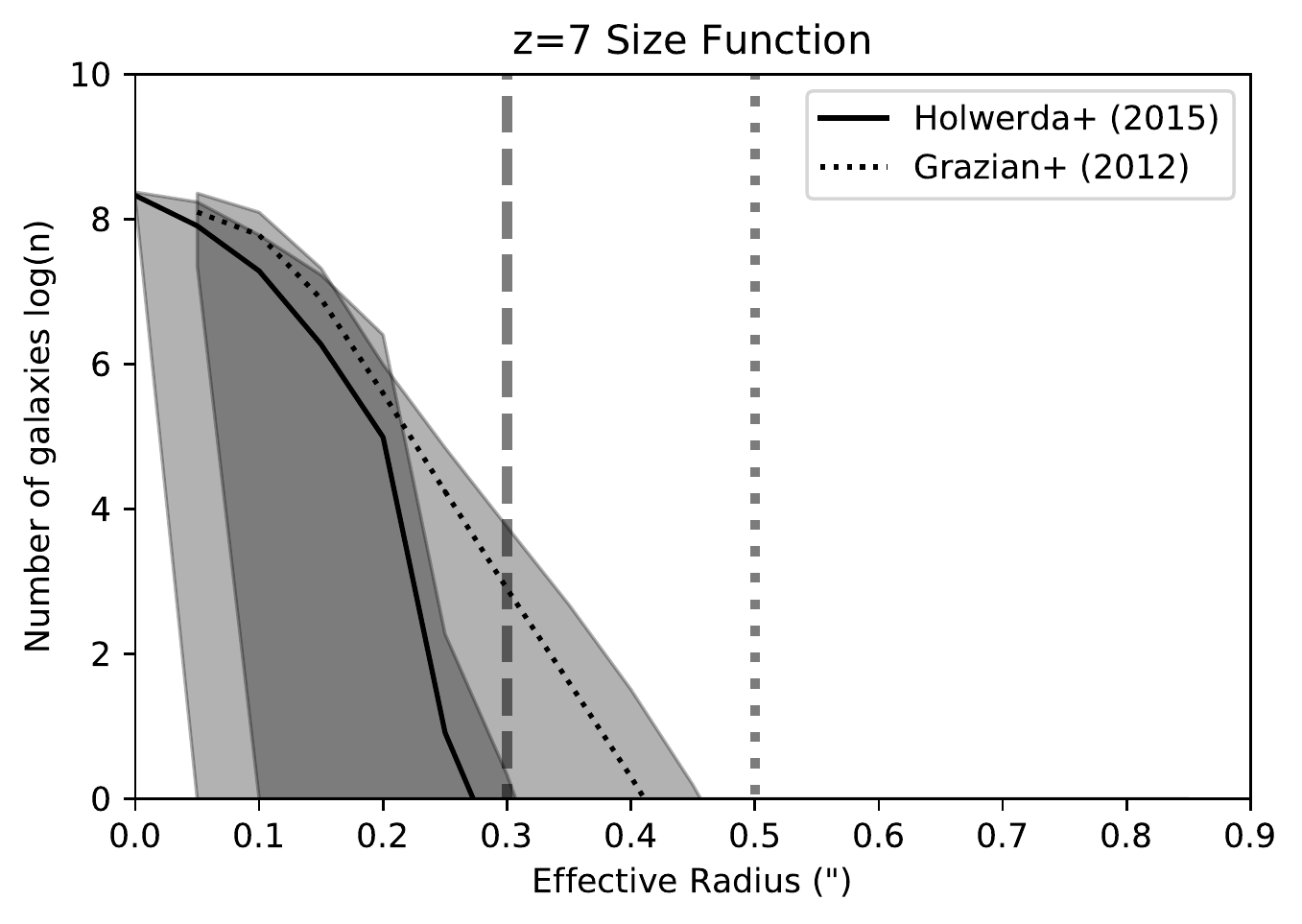}
\includegraphics[width=0.49\textwidth]{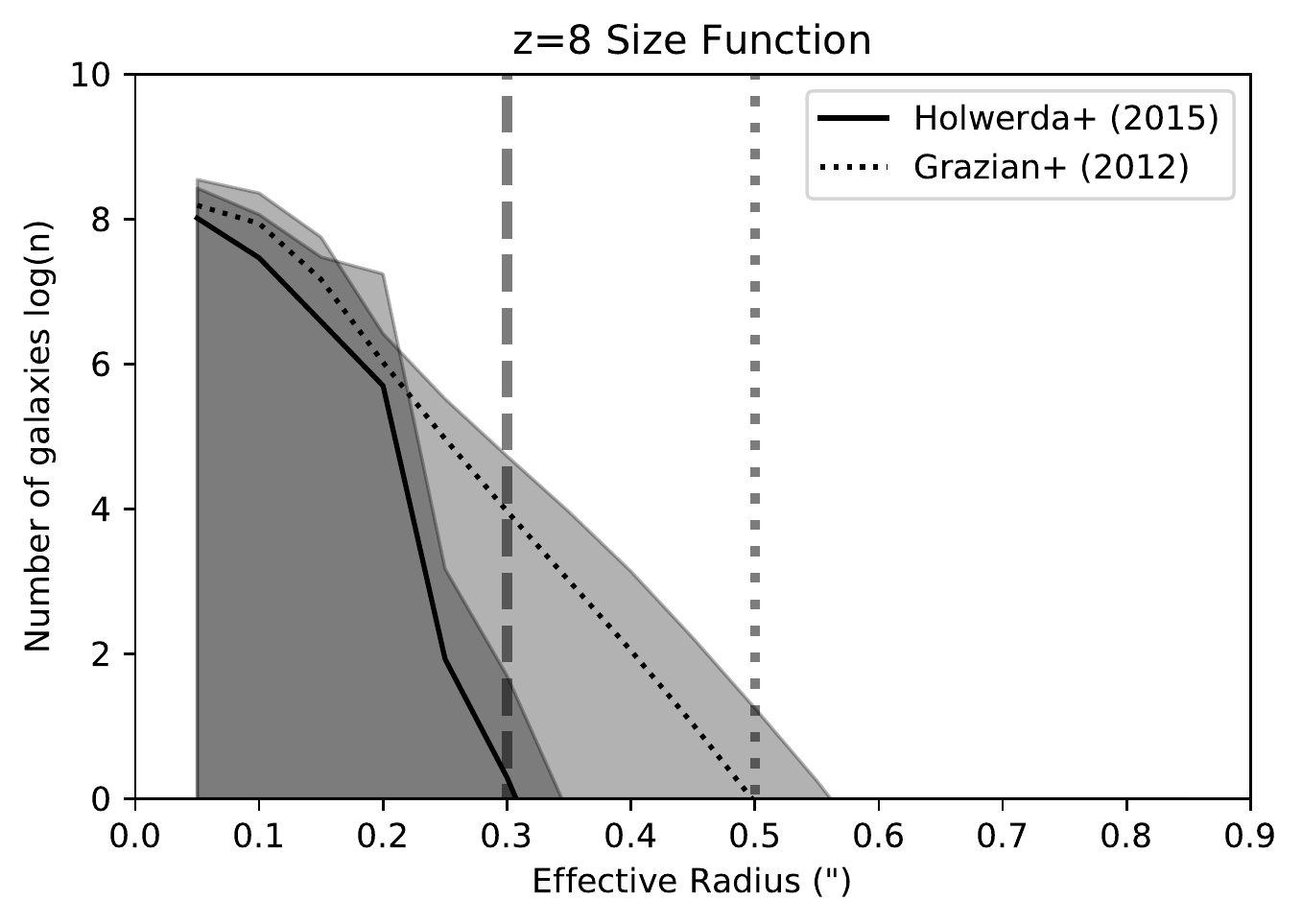}
\includegraphics[width=0.49\textwidth]{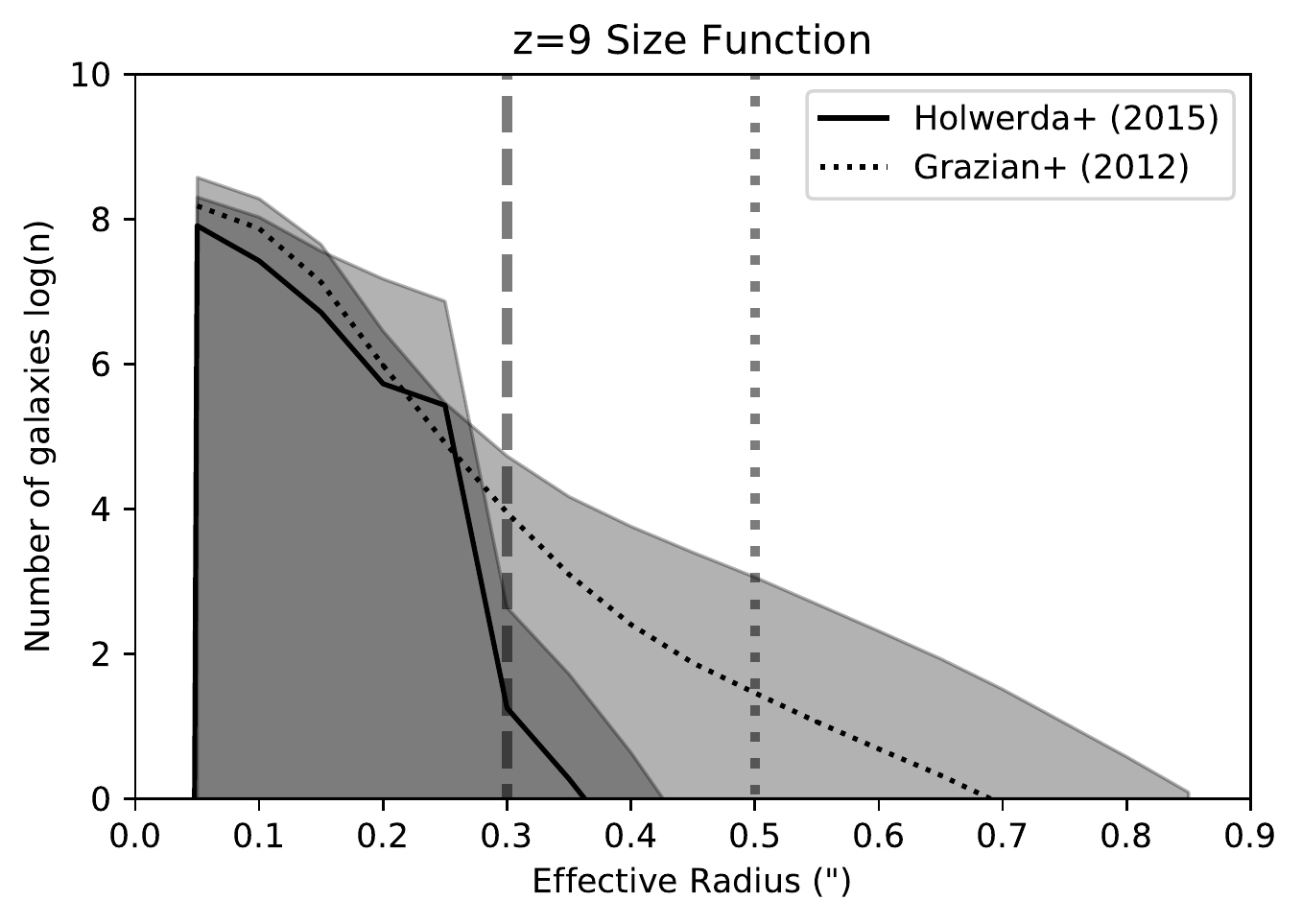}
\includegraphics[width=0.49\textwidth]{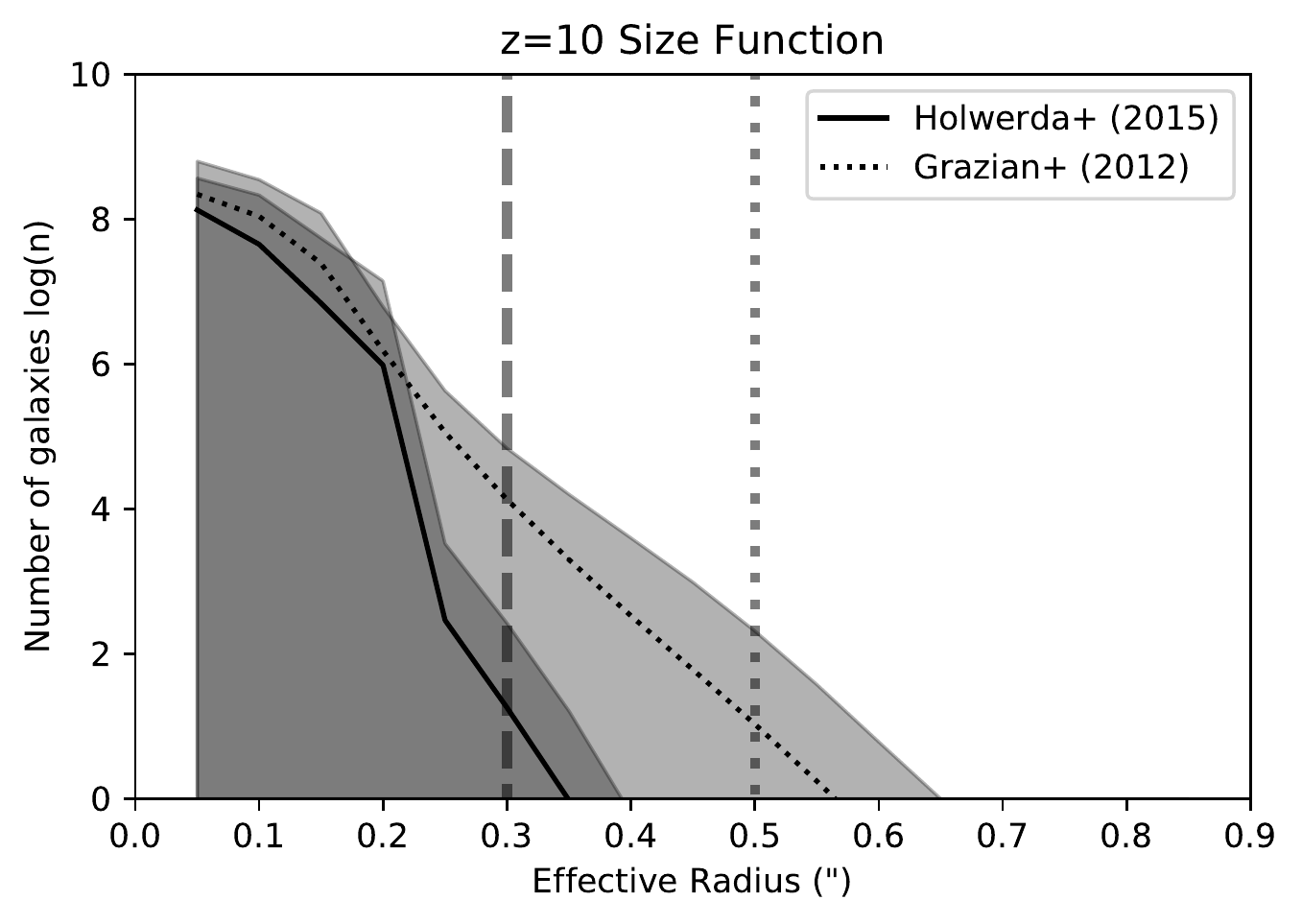}
\includegraphics[width=0.49\textwidth]{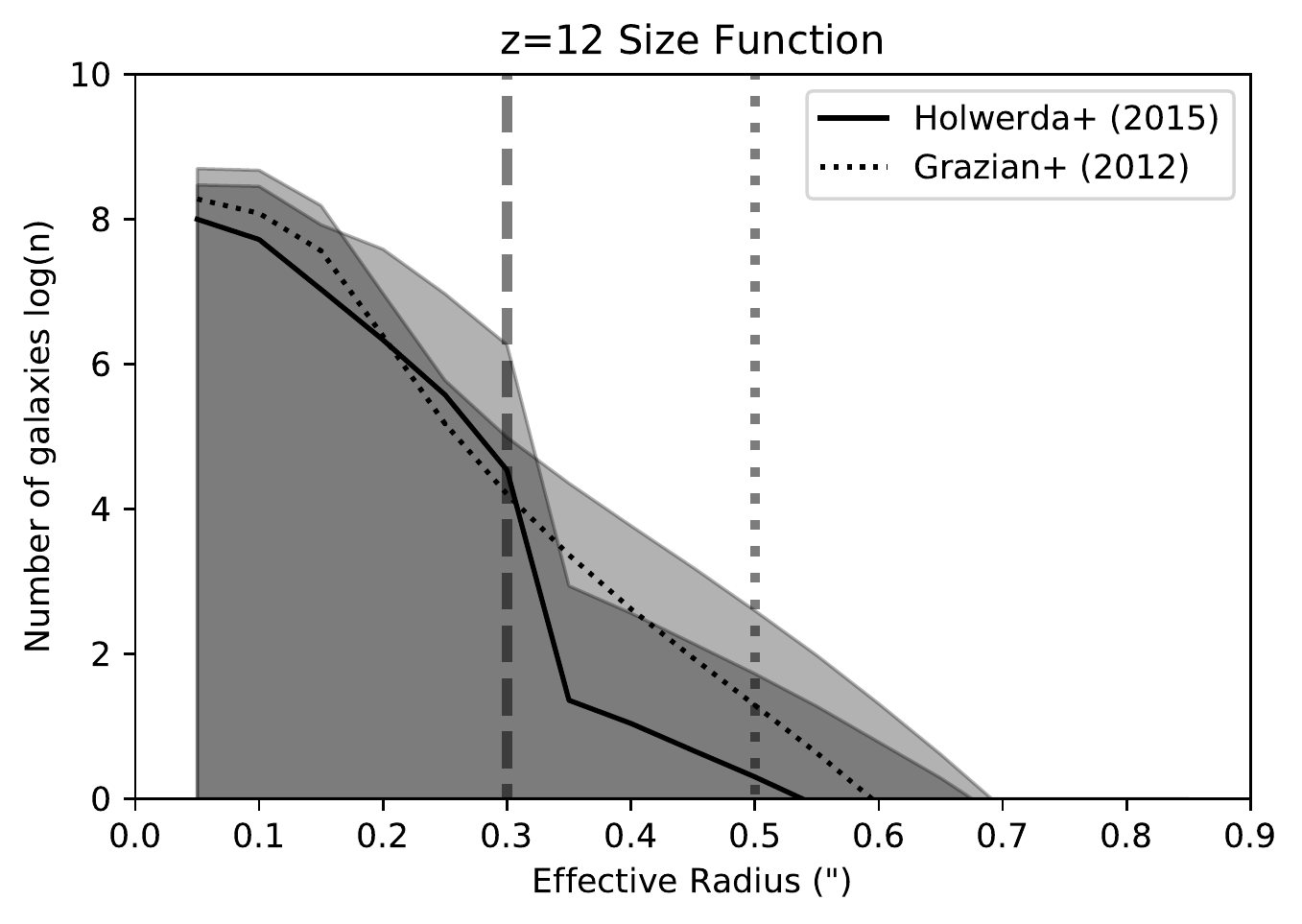}
\includegraphics[width=0.49\textwidth]{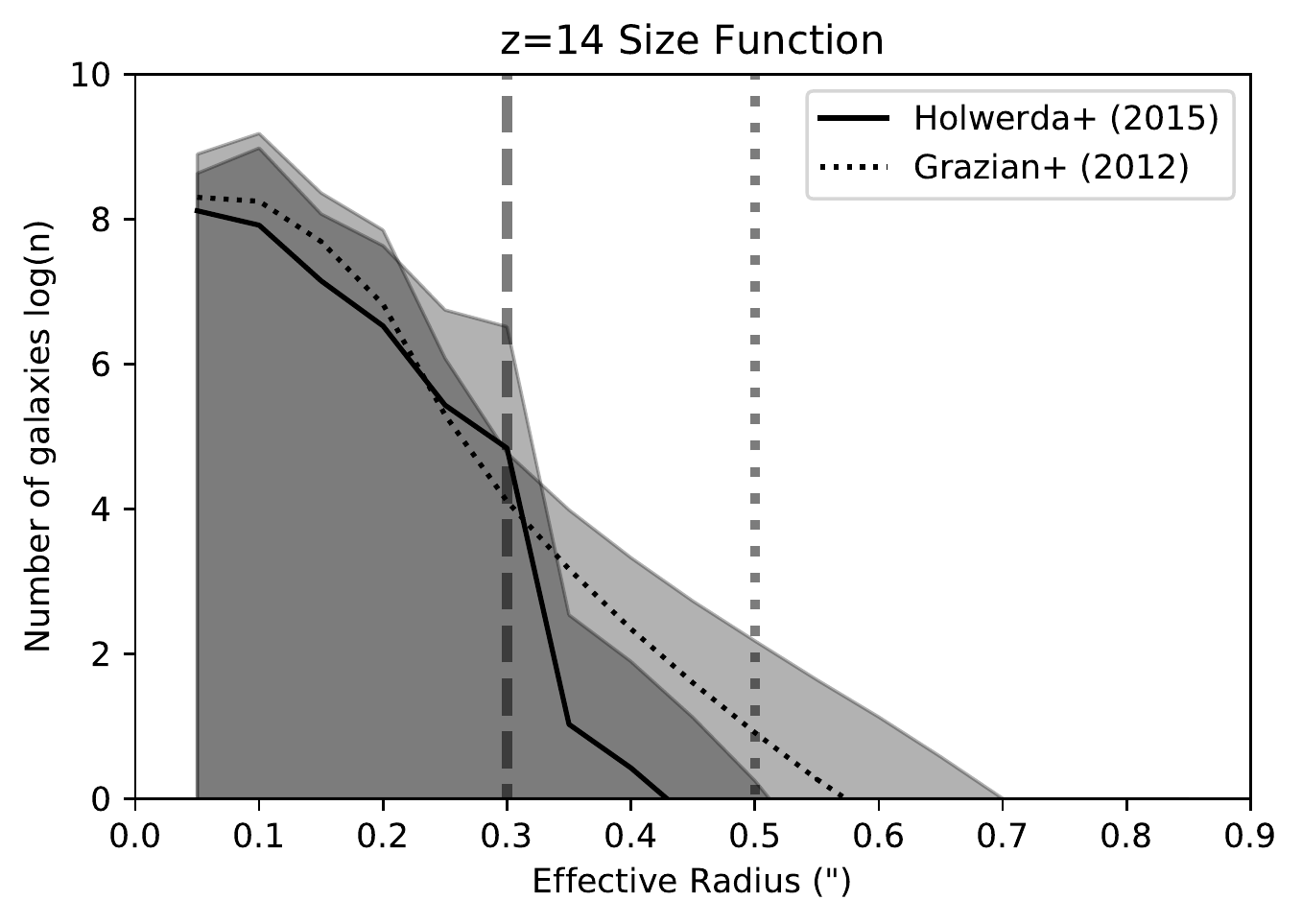}

\caption{The mean and standard deviation variance of the size functions assuming the luminosity-size relation from either \cite{Grazian12} (dashed) or the \cite{Holwerda15} (solid) for the z=7--14 epochs. Uncertainty was bootstrapped using the \cite{Mason15} luminosity function uncertainties and those in each luminosity-size relation. }
\label{f:sizefunctions}
\end{figure*}

Examples of the resulting size functions at z=7,8 and 9 are in Figure \ref{f:sizefunctions}. The evolution with redshift is noticeable but the difference in the assumed size-luminosity relation is driving the fraction of objects removed with a size cut.  
The mean and standard deviation of the fraction of galaxies rejected by a 0\farcs3 or 0\farcs5 size cut are listed in Table \ref{t:lossrate}. We note that the uncertainty in the loss rate is much greater than the mean, principally because the uncertainties in the luminosity-size relations. 

A hard size cut would not result in a significant loss of sample in the case of the \cite{Holwerda15} relation but it could be as large as the contamination rate of bright sources if one assumes the \cite{Grazian12} relation. And since these loss rates are for individual sources, close pairs or mergers might be excluded altogether, which is also an issue for visual inspection. With that in mind, using size as a Baysian prior for high redshift candidacy (flat to 0\farcs3 and dropping zero at 0\farcs5).

\begin{table*}[htp]
\caption{The loss rate percentage --the percentage of all detectable galaxies at each redshift removed-- of the 0\farcs3 and 0\farcs5 hard cut assuming one of the two different size-luminosity relations and the \cite{Mason15} luminosity functions over the observable luminosity range for BoRG[z9].
The error predominantly comes from uncertainty in the size-luminosity relation and the size measurement, assumed to be a WFC3/IR FWHM (0\farcs1). }
\begin{center}
\begin{tabular}{l l l }
z & \cite{Holwerda15}	& \cite{Grazian12}	\\
 & (\%)	& (\%)		\\
 \hline
 \hline
5.00 & 0.00$\pm$0.00 & 15.26$\pm$28.74 \\
6.00 & 0.00$\pm$0.02 & 25.86$\pm$34.04 \\
7.00 & 0.01$\pm$0.07 & 47.27$\pm$44.69 \\
8.00 & 2.03$\pm$14.00 & 72.10$\pm$40.01 \\
9.00 & 3.05$\pm$17.06 & 88.40$\pm$28.69 \\
10.00 & 2.80$\pm$15.96 & 95.85$\pm$17.69 \\
12.00 & 13.02$\pm$32.85 & 99.96$\pm$0.41 \\
14.00 & 34.08$\pm$47.32 & 99.76$\pm$2.38 \\
16.00 & 48.45$\pm$48.38 & 100.00$\pm$0.00 \\
\hline
\hline
5.00 & 0.00$\pm$0.00 & 10.56$\pm$22.82 \\
6.00 & 0.00$\pm$0.00 & 12.15$\pm$24.88 \\
7.00 & 0.00$\pm$0.00 & 11.34$\pm$24.82 \\
8.00 & 0.00$\pm$0.00 & 15.93$\pm$27.75 \\
9.00 & 0.00$\pm$0.00 & 22.45$\pm$34.60 \\
10.00 & 0.00$\pm$0.00 & 28.50$\pm$38.84 \\
12.00 & 0.00$\pm$0.00 & 47.26$\pm$45.16 \\
14.00 & 2.00$\pm$14.00 & 77.23$\pm$37.32 \\
16.00 & 3.03$\pm$17.06 & 89.23$\pm$29.67 \\
\hline
\end{tabular}
\end{center}
\label{t:lossrate}
\end{table*}%

\end{document}